\documentclass{icrc2009}

\usepackage{graphicx}   
\usepackage[caption=false]{caption}    
\usepackage[font=footnotesize]{subfig} 
\usepackage{fixltx2e}
\usepackage{url}

\newcommand{\shorttitle}[1]%
{\markboth{Proceedings of the 31\MakeLowercase{$^{st}$} ICRC, {\L}\'{o}d\'{z} 2009}{#1} }
\newcommand{\etal}{\MakeLowercase{\textit{et al. }}} 

\begin{document}

\title{Search for the Kaluza-Klein Dark Matter with the AMANDA/IceCube Detectors}

\author{\IEEEauthorblockN{Matthias Danninger\IEEEauthorrefmark{1}
			  and Kahae Han\IEEEauthorrefmark{2}
                          for the IceCube Collaboration\IEEEauthorrefmark{3}}
                            \\
\IEEEauthorblockA{\IEEEauthorrefmark{1}Department of Physics, Stockholm University, AlbaNova, S-10691 Stockholm, Sweden}
\IEEEauthorblockA{\IEEEauthorrefmark{2}Department of Physics and Astronomy, University of Canterbury, Pr. Bag 4800 Christchurch, New Zealand}
\IEEEauthorblockA{\IEEEauthorrefmark{3}See http://www.icecube.wisc.edu/collaboration/authorlists/2009/4.html}}

\shorttitle{M. Danninger \etal Kaluza-Klein Dark Matter Detection in AMANDA/IceCube}
\maketitle

\begin{abstract}
  A viable WIMP candidate, the lightest Kaluza-Klein particle (LKP),
  is motivated by theories of universal extra dimensions.
  LKPs can scatter off nuclei in large celestial bodies, like the Sun, and become trapped within their deep 
  gravitational wells, leading to high WIMP densities in the object's core. Pair-wise 
  LKP annihilation could lead to a detectable high energy neutrino flux 
  from the center of the Sun in the IceCube neutrino telescope.
  
  We describe an ongoing search for Kaluza-Klein solar WIMPs with the AMANDA-II 
  data for the year 2001, and also present a UED dark matter sensitivity projected 
  to 180 days from a study of data taken with the combined AMANDA II and IceCube detector 
  in the year 2007. A competitive sensitivity, compared to existing direct and indirect 
  search experiments, on the spin-dependent cross section of the LKP on 
  protons is also presented.
\end{abstract}

 
\section{Introduction}
\begin{figure}[!t]
  \centering
  \includegraphics[width=2.7in]{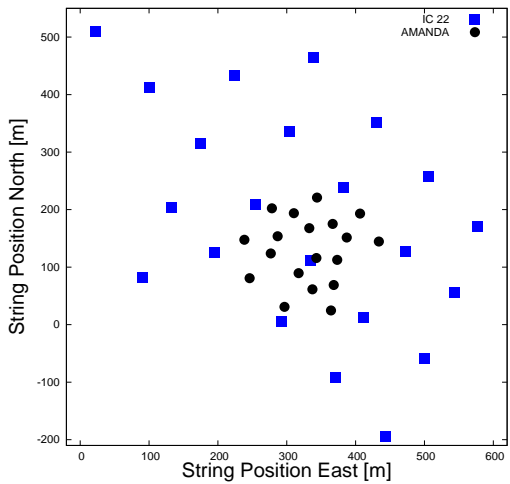}
  \caption{Top view of the 2007 IceCube+AMANDA detector configuration. The IceCube-22 strings (squares) enclose the AMANDA strings (circles).}
  \label{fig1}
\end{figure}

Kaluza-Klein weakly interacting massive particles (WIMP) arising from theories with extra dimensions have come under increased scrutiny \cite{ref6} alongside WIMP candidates from supersymmetric particle theories, e.g. the neutralino.\\Several analyses \cite{ref1,ref2} performed on the data from the AMANDA-II and the IceCube detectors have already put limits on the neutralino induced muon flux from the Sun comparable to that of direct detection experiments.
The first excitation of the Kaluza Klein (KK) photon, $B^{(1)}$, in the case of Universal Extra Dimensions (UED) with one extra dimension, annihilates to all standard model particles. This results in the production of a detectable flux of muon neutrinos in the IceCube detector. $B^{(1)}$ is often referred to as the LKP - lightest Kaluza-Klein Particle. KK-momentum conservation leads to the stability of the LKP, which makes it a viable dark matter candidate.
Compared to neutralino WIMPs, LKPs come from a relatively simple extension of the Standard Model and, consequently, branching ratios (see Table I) and cross sections are calculated with fewer assumptions and parameter-dependences.
\begin{table}[!t]
 \caption{Possible channels for the pair annihilation of $B^{(1)}B^{(1)}$ and branching ratios of the final states. Figures taken from \cite{ref19}.}
 \centering
 \begin{tabular}{c @{$\rightarrow$}l | c} 
  \multicolumn{2}{c|}{Annihilation Process} & Branching ratio \\ 
  \hline 
  $B^{(1)}B^{(1)}\quad$ &  $\quad \nu_{e} \overline{\nu}_{e}$, $\nu_{\mu} \overline{\nu}_{\mu}$, $\nu_{\tau} \overline{\nu}_{\tau}$ &  $0.012$ \\
                  &  $\quad e^{+}e^{-}$, $\mu^{+}\mu^{-}$, $\tau^{+}\tau^{-}$ &  $0.20$  \\
		  &  $\quad u\overline{u}$, $c\overline{c}$, $t\overline{t}$  &  $0.11$  \\
		  &  $\quad d\overline{d}$, $s\overline{s}$, $b\overline{b}$  &  $0.007$  \\
 \end{tabular}
 \label{table_1}
\end{table}
This feature allows us to perform a combined channel analysis for an LKP particle.
Another consequence of the simple UED model is that with the assumption of a compactified extra dimension scale of around 1TeV, the particle takes a much narrower range of masses \cite{ref6} from the relic density calculation - ranging from $600$ GeV to $800$ GeV and $500$ GeV to $1500$ GeV if coannihilations are accounted for \cite{ref11}. Moreover, collider search limits rule out LKP masses below $300$ GeV \cite{ref14,ref15}.\\
In this paper we describe an ongoing solar WIMP analysis with the (2001) AMANDA data. Furthermore, we derive for the combined geometry of $22$ IceCube strings (IC22) and AMANDA (to be referred to as the combined analysis in the rest of the paper) the projected sensitivity on the muon flux and spin-dependent (SD) cross section obtained for LKP WIMPs with data from the year 2007.\\
The AMANDA-II detector, a smaller predecessor of IceCube with $677$ OMs on $19$ strings, ordered in a $500$m by $200$m diameter cylindrical lattice, has been fully operational since 2001 \cite{refAmanda}. The IceCube Detector, with its $59^{th}$ string deployed this season, is much larger with increased spacing between the strings and will have a total instrumented volume of $1$km$^{3}$ \cite{ref20}. The set-up in 2007 for the combined analysis consisted of $22$ IceCube strings, and the $19$ AMANDA strings, with a separate trigger and data acquisition system. The detector geometry for both AMANDA-II and IC22 is shown in Fig. \ref{fig1}.
\section{Simulations}
A solar WIMP analysis can be thought of as using the Earth as its primary physical filter for data, as one only looks at data collected when the Sun is below the horizon at the South Pole, $\Theta_{\odot}\in[90^{\circ},113^{\circ}]$. Single\footnote{atmospheric muons from single CR showers}, $\mu_{\it{single}}$, and coincident\footnote{atmospheric muons from coincident CR showers}, $\mu_{\it{coin}}$, atmospheric muons that come from cosmic ray showers in a zenith angle range $\Theta_{\mu}$ of $[0^{\circ},90^{\circ}]$, constitute the majority of the background, whereas the near-isotropic distribution of atmospheric neutrinos, $\nu_{\it{atm}}$, will form an irreducible background. The atmospheric muon backgrounds are generated using CORSIKA \cite{ref5} with the H\"{o}randel CR composition model \cite{ref12}. For the atmospheric neutrino background, produced according to the Bartol model \cite{ref13}, ANIS \cite{ref8} is used.
For the combined analysis, the simulated $\mu_{\it{single}}$ background has a detector-livetime of $1.2$ days, $\mu_{\it{coin}}$ of $7.1$ days and $\nu_{\it{atm}}$ of $9.8$ years.
\begin{figure*}[!t]
   \centerline{\subfloat[Effective volume]{\includegraphics[width=2.9in]{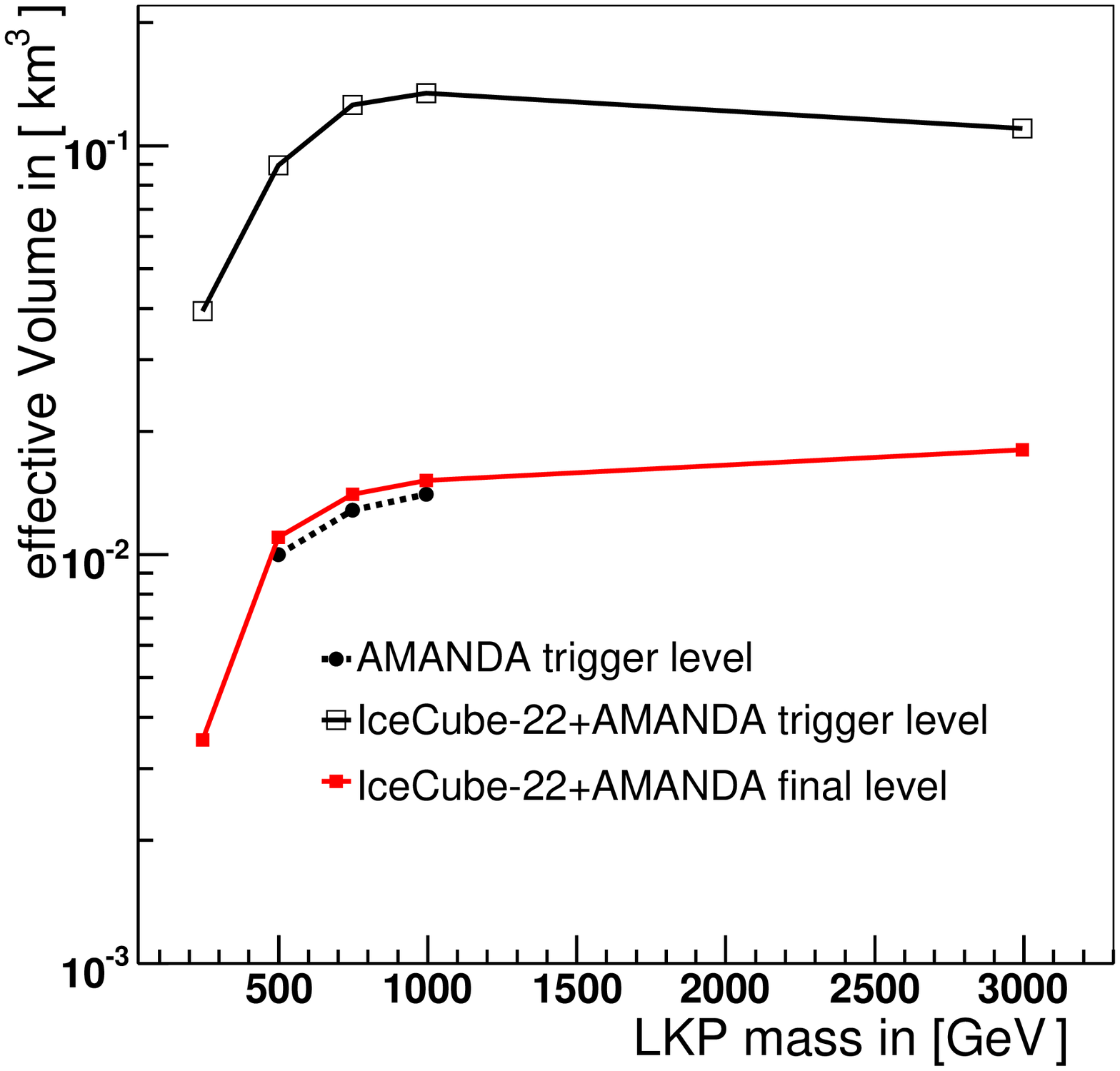} \label{fig2}}
              \hfil
              \subfloat[Muon flux sensitivity]{\includegraphics[width=2.9in]{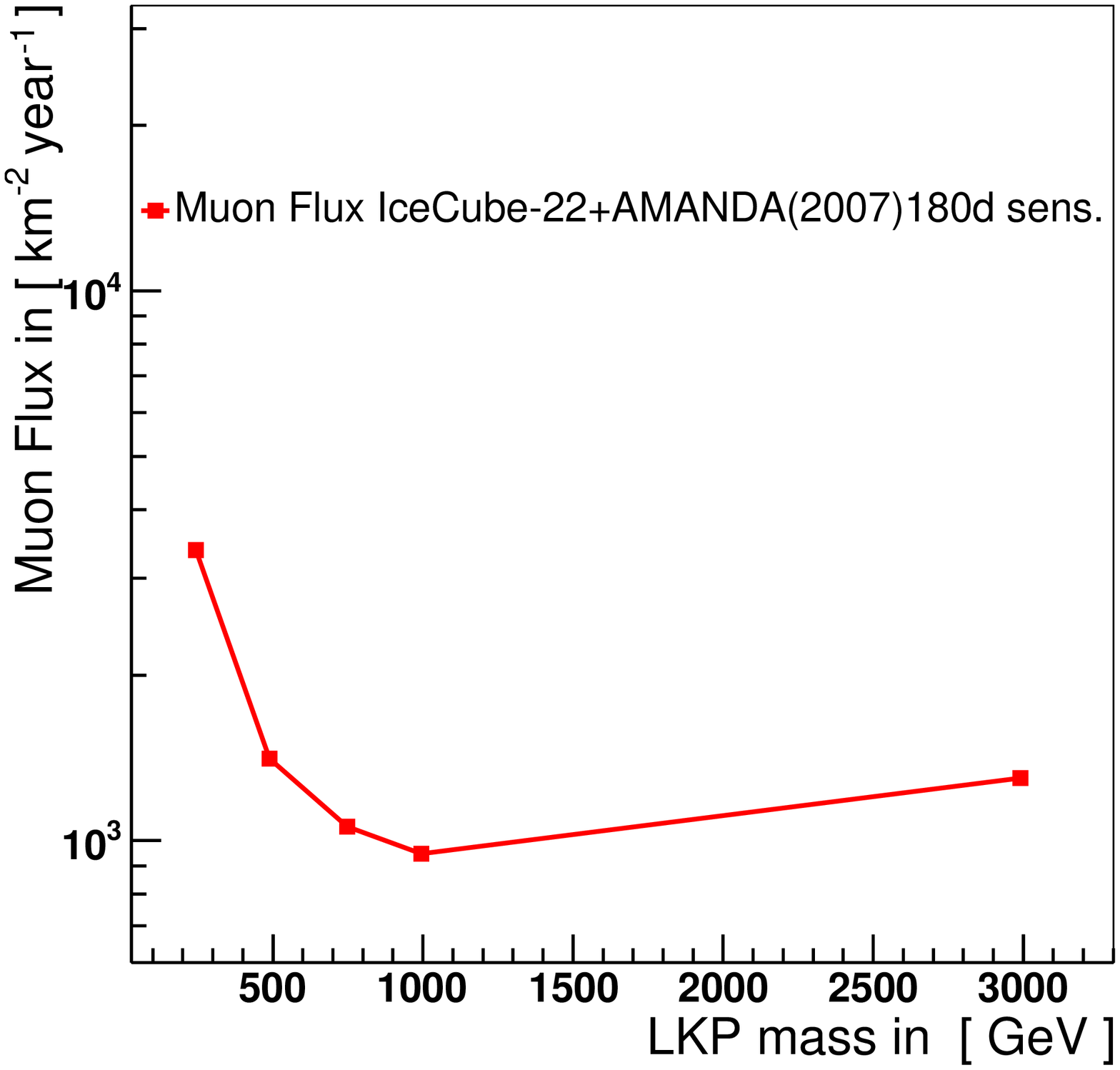} \label{fig3}}
             }
   \caption{Fig.2a shows the effective volume as a function of LKP mass at trigger level and final cut level for the IceCube-22+AMANDA analysis and at trigger level only for the AMANDA analysis. Fig.2b demonstrates the projected sensitivity to $180$ days of livetime on the muon flux from LKP annihilations in the Sun as a function of LKP mass for the IceCube-22+AMANDA detector configuration.}
   \label{double_fig}
\end{figure*}
WIMPSIM \cite{ref3,ref4} was used to generate the signal samples for LKP WIMPs, consisting of $2$ million events per channel for WIMP masses varying from $250$ GeV to $3000$ GeV. Individual annihilation channels (three $\nu$'s, $\tau$, and t,b,c quarks), contributing to $\nu_{\mu}$'s at the detector, were generated for the combined analysis, as well as for the AMANDA only analysis (in the latter case for the energy range from $500$ GeV to $1000$ GeV).
Muon and \v{C}erenkov light propagation in Antarctic ice were simulated using IceCube/AMANDA software such as MMC \cite{ref7}, PTD and photonics \cite{ref77}. Finally, AMASIM for AMANDA and ICESIM for the combined detector were used to simulate the detector response.
The signal detection efficiency of the two detector configurations is given by the effective volume, $V_{\it{eff}}$, which is defined for a constant generation volume, $V_{\it{gen}}$, by
\begin{equation}\label{equ1}
V_{\it{eff}}=V_{\it{gen}}\cdot\frac{N_{\it{obs}}}{N_{\it{gen}}} ,
\end{equation}
where $N_{\it{obs}}$ is the number of observed LKP events and $N_{\it{gen}}$ the number of generated LKP events, undergoing charged-current interaction within $V_{\it{gen}}$. $V_{\it{eff}}$ is a good quantity to compare LKP detectability at trigger level for the two analyses, shown in Fig. \ref{fig2}.\\
After deadtime correction, 142.5 days of data when the Sun was below the horizon were available in 2001 with a total number of $1.46\cdot 10^{9}$ recorded events for the AMANDA analysis.
The combined analysis is utilizing a projected total livetime of $180$ days of data for the calculated sensitivities in this paper.\\
\\The main purpose of the Monte Carlo (MC) simulations of the various background sources is to show that a good agreement with experiment is achieved, demonstrating a sufficient understanding of the detector. Thus, it is viable to assume that the LKP signal samples are simulated correctly within the AMANDA/IceCube simulation-chain and can be used to select the different cut parameters for the higher cut levels $L2, \ L3 \ \mbox{and} \ L4$, because their difference from background in different parameter distributions can be clearly identified. The actual cut value of each cut level is obtained by maximizing the efficiency function, or a figure-of-merit, for simulated LKP signals and the experimental background sample, which consists of data taken when the Sun was above the horizon and therefore contains no solar WIMP signal. Setting cut values based on experimental background datasets has the advantage that possible simulation flaws are minimized.
\section{Filtering}
LKP signals are point sources with very distinct directional limitations (zenith angle theta, $\Theta_{\it{zen}}=90^{\circ}\pm23^{\circ}$). Hence, the general strategy of filtering for both analyses is to apply strict directional cuts in early filter levels.
$L0$ and $L1$ consist of calibration, reconstruction and making a simple angular cut of $\Theta_{\it{zen}}>70^{\circ}$ on the first-guess reconstructed track. This leads to a passing efficiency of around $0.7$ for all LKP signal samples, and reduction of around $0.002$ for both, data and muon background. All events passing the $L0+L1$ level are reconstructed using log-likelihood methods (llh). $L2$ is a two dimensional cut on the reconstructed llh-fit zenith angle ($\Theta_{zen,llh}$) within $\Theta_{\odot}$ and the estimated angular uncertainty of the llh track.
$L3$ picks reconstructed tracks, which are nearly horizontal and pass the detector, to further minimize vertical tracks associated with background events. The multivariate filter level, $L4$, consists of two different multivariate analysis routines from the TMVA \cite{ref9} toolkit, namely a support vector machine (SVM) together with a Gaussian fit-function and a neural network (NN). The input variables for the two algorithms are obtained by choosing parameters with low correlation but high discrimination power between background and signal. 
The individual output parameters are combined in one multivariate cut parameter $Q_{\it{NN}}\cdot Q_{\it{SVM}}$.\\
\section{Sensitivity}
After the $L4$ cut\footnote{starting with $L4$, only the combined analysis is discussed}, the muon background reduction is better than a factor $1.16\cdot 10^{-7}$, which implies that the final sample is dominated by $\nu_{\it{atm}}$ background. The solar search looks for an excess in neutrino events over the expected background in a specifically determined search cone towards the direction of the Sun with an opening angle $\Psi$. Events with a reconstructed track
direction pointing back towards the Sun within an angle $\Psi$ are kept,
where $\Psi$ is optimized to discriminate between the $\nu_{\it atm}$
background and a sum of all seven LKP channels, weighted with the
expected branching ratios as listed in Table I.\\
The expected upper limit, or sensitivity, for an expected number of background events $n_{\it{Bg}}$ is
\begin{equation}
 \overline{\mu}_{s}^{90\%}(n_{\it{Bg}}) = \sum_{n_{\it{obs}=0}}^{\infty}\mu_{s}^{90\%}(n_{\it{obs}})\frac{(n_{\it{Bg}})^{n_{\it{obs}}}}{(n_{\it{obs}})!}e^{-n_{\it{Bg}}}\ ,
 \label{solarEQ3}
\end{equation}
where $\mu_{s}^{90\%}(n_{obs})$ is the Feldman-Cousins upper limit for the number of observed events, $n_{\it{obs}}$ \cite{ref17}.
The model rejection factor \cite{ref18}
\begin{equation}
 \it{MRF} = \frac{\overline{\mu}_{s}^{90\%}}{n_{s}} ,
 \label{SunOptimEQ}
\end{equation}
is used to determine the optimum opening angle $\Psi$ of the solar search cone. Here, $n_{s}$ is the number of surviving LKP events within $\Psi$.\\
\\
Under the assumption of no signal detection, it is possible to derive the Feldman-Cousins sensitivity discussed above for the combined detector with a total projected livetime of $T_{\it{live}}=180$ days. The expected number of events after cut level $L4$ are estimated from a processed subset of observational data with a detector livetime of $5.61$ days. The results are then extrapolated to the total livetime $T_{\it{live}}$, yielding an expectation of $7140$ events. The corresponding expectation from the simulated background samples, $n_{\it{Bg,MC}}$, normalized to the data at filter level $L1$ and extrapolated to $T_{\it{live}}$, is $633(\mu_{\it{coin}})+1038(\mu_{\it{single}})+5340(\nu_{\it{atm}})=7011(n_{\it{Bg,MC}})$.\\
The expected sensitivity on the neutrino-to-muon conversion rate $\overline{\Gamma}_{\nu \rightarrow \mu}^{90\%}$ is given by
\begin{equation}
 \overline{\Gamma}_{\nu \rightarrow \mu}^{90\%} = \frac{\overline{\mu}_{s}^{90\%}}{V_{\it{eff}}\cdot T_{\it{live}}} \ ,
 \label{SensitivityEQ1}
\end{equation}
where the effective volume $V_{\it{eff}}$ is given by eq. \ref{equ1}. For each annihilation channel, one can separately calculate the $V_{\it{eff}}$ within the solar search cone, determined by the combined signal p.d.f., $f_{S}^{\it{all}}(x|\Psi)$, and thereby determine a $\overline{\Gamma}_{\nu \rightarrow \mu}^{90\%}$ for each channel. Additionally, the combined effective volume, $V_{\it{eff,LKP}}$, for the expected $\nu_{\it{LKP}}$ spectrum is given by the sum of the individual $V_{\it{eff}}$ per channel, weighted with the respective branching ratio of each channel. For the neutrino-to-muon conversion rate per single channel, the expected limit on the annihilation rate in the core of the Sun per second is given by, 
\begin{equation}
 \overline{\Gamma}_{A}^{90\%} = (c_{1}(\it{ch},m_{B^{(1)}}))^{-1}\cdot \overline{\Gamma}_{\nu \rightarrow \mu}^{90\%} \ , 
 \label{SensitivityEQ2}
\end{equation}
where $c_{1}(\it{ch},m_{B^{(1)}})$ is an LKP annihilation channel ($\it{ch}$) and energy dependent constant. The sensitivity to the muon flux at a plane at the combined detector is derived via the calculation chain $\overline{\Gamma}_{\nu \rightarrow \mu}^{90\%}\rightarrow \overline{\Gamma}_{A}^{90\%}\rightarrow \Phi^{90\%}_{\mu}$ and is performed using the code described in \cite{ref3,ref4}.
\begin{figure*}[th]
  \centering
  \includegraphics[width=4.4in]{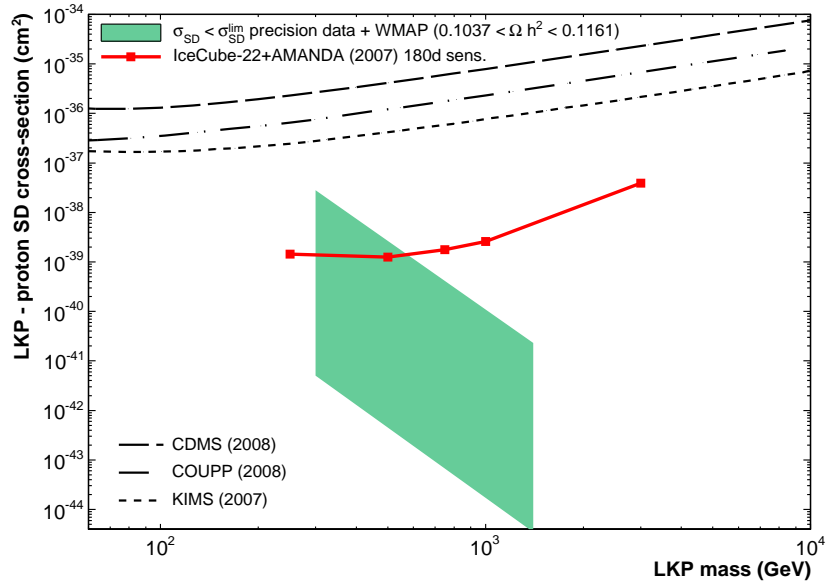}
  \caption{Theoretically predicted spin-dependent $B^{(1)}$-on-proton elastic scattering cross sections are indicated by the shaded area \cite{ref16}. The cross-section prediction vary with the assumed mass of the first KK excitation of the quark, constrained by $0.01\leq r = (m_{q^{(1)}}-m_{B^{(1)}})/m_{B^{(1)}}\leq0.5$. The current `best' limits, set by direct search experiments are plotted together with the sensitivity of the combined detector IceCube-22+AMANDA. The region below $m_{B^{(1)}}=300$ GeV is excluded by collider experiments \cite{ref14,ref15} and $m_{B^{(1)}}>1500$ GeV is strongly disfavored by WMAP observations \cite{ref10}.
    }
  \label{fig4}
\end{figure*}
The results for the final $V_{\it{eff}}$ and the predicted sensitivity to a muon flux resulting from LKP induced annihilations in the Sun for the combined IC22 and AMANDA detector 2007 with a total livetime of $180$ days are presented in figures \ref{fig2} and \ref{fig3}. From the derived $\nu$-to-$\mu$ conversion rate, $\overline{\Gamma}_{\nu \rightarrow \mu,\it{LKP}}^{90\%}$, we can calculate the sensitivity for the annihilation rate in the Sun per second, $\overline{\Gamma}_{A,\it{LKP}}^{90\%}$.\\
In \cite{ref19}, it is shown that the equilibrium condition between $\Gamma_{A,\it{LKP}}$ and the capture rate $C^{\odot}$ is met by LKPs within the probed mass range. Furthermore, the capture rate of LKPs in the Sun is entirely dominated by the spin-dependent component of the $B^{(1)}$-on-proton elastic scattering \cite{ref21}. Consequently, presuming an equilibrium of $\Gamma_{A,LKP}=C^{\odot}$, the sensitivity for the spin-dependent elastic scattering cross section\footnote{The local density of DM in our galaxy is taken to match the mean density $\overline{\rho}_{DM}=0.3$ GeV/c$^{2}$cm$^{3}$, and the rms velocity is set to $\overline{v}=270$ km/s.} of $B^{(1)}$ can be calculated as,   
\begin{equation}
  \sigma_{\it{H,SD}} \simeq 0.597 \cdot 10^{-24}\rm{pb} \left(\frac{m_{B^{(1)}}}{1 \it{TeV}} \right)^{2}\cdot \Bigg(\frac{\overline{\Gamma}_{\it{A,LKP}}^{90\%}}{s^{-1}}\Bigg) .
 \label{somemoreEq}
\end{equation}
The estimated sensitivity for the spin-dependent cross section for LKPs is displayed in figure \ref{fig4}, along with the most recently published limits from direct search experiments. The theoretical spin-dependent cross section predictions (shaded area) for LKPs are taken from \cite{ref16} and are plotted for different predictions for the mass of the first KK-excitation of the quark.
\section{Conclusion and Outlook}
We showed that a competitive result on the spin-dependent cross-section of LKP-on-proton scattering can be obtained with the combined geometry of AMANDA-II and IceCube-22, which explores parts of the unrejected regions in the theoretically predicted LKP-region.\\ 
\\We also described the ongoing solar WIMP analysis of the AMANDA-II data taken during 2001. This will be extended to include 2002 and 2003 data. Furthermore, as the energy signature of $\nu_{\mu}$'s induced by LKP annihilations in the Sun is very hard, the fullsized IceCube-80 detector will markedly improve the sensitivity and set strong limits on LKP WIMP theories.

\end{document}